# Bandstructure Effects in Silicon Nanowire Hole Transport

Neophytos Neophytou, Abhijeet Paul, and Gerhard Klimeck

School of Electrical and Computer Engineering, Purdue University, West Lafayette, Indiana 47907-1285

Email: neophyto@purdue.edu

## ABSTRACT

Bandstructure effects in PMOS transport of strongly quantized silicon nanowire field-effect-transistors (FET) in various transport orientations are examined. A 20-band $sp^3d^5s^*$ spin-orbit-coupled (SO) atomistic tight-binding model coupled to a self consistent Poisson solver is used for the valence band dispersion calculation. A ballistic FET model is used to evaluate the capacitance and current-voltage characteristics. The dispersion shapes and curvatures are strong functions of device size, lattice orientation, and bias, and cannot be described within the effective mass approximation. The anisotropy of the confinement mass in the different quantization directions can cause the charge to preferably accumulate in the (110) and secondly on the (112) rather than (100) surfaces, leading to significant charge distributions for different wire orientations. The total gate capacitance of the nanowire FET devices is, however, very similar for all wires in all the transport orientations investigated ([100], [110], [111]), and is degraded from the oxide capacitance by ~30%. The [111] and secondly the [110] oriented nanowires indicate highest carrier velocities and better ON-current performance compared to [100] wires. The dispersion features and quantization behavior, although a complicated function of physical and electrostatic confinement, can be explained at first order by looking at the anisotropic shape of the heavy-hole valence band.

**Index terms – nanowire, bandstructure, tight binding, transistors, PMOS, hole, valence band, MOSFETs, non-parabolicity, effective mass, injection velocity, quantum capacitance, anisotropy.**

# 1. Introduction

*Motivation:* As transistor sizes shrink down to the nanoscale, existing CMOS field-effect transistors are expected to evolve from planar, to 3D non-planar devices at nanometer sizes [1]. The use of nanowire (NW) transistors as field-effect-transistor (FET) devices has recently attracted large attention, because such structure could provide enhanced electrostatic control. Nanowire transistors of diameters even down to 3nm have already been demonstrated by various experimental groups [2, 3, 4, 5, 6]. This work investigates hole transport properties of nanowire devices by using a nearest-neighbor 20 band tight-binding (TB) model ($sp^3d^5s^*$ with spin-orbit-coupling) [7, 8, 9, 10] for the electronic structure calculation, coupled to a 2D Poisson solver for electrostatics. It is a continuation of the work presented in [11] that investigates bandstructure effects in electron transport of n-type nanowires. To evaluate transport characteristics, a simple semi-classical ballistic model [12, 13] (Fig. 1a) is used. The ballistic transport characteristics of square p-type nanowires of 3nm and 6nm sides, oriented in [100], [110] and [111] transport directions are examined and compared.

*Necessity of atomistic modeling for the valence band:* Under extreme scaling of the device's dimensions, the atoms in the cross section will be countable, and crystal symmetry, bond orientation and quantum mechanical confinement will be critical. The problem of identifying the correct bandstructure for the valence band of Si in the inversion layers is much more complicated than the corresponding conduction band counterpart because of the strong non-parabolicity and anisotropy of the heavy-hole, and the coupling to the light-hole and the split-off band. *k.p* methods have been traditionally used [14, 15] for both un-strained and strained channels. De Michielis recently in [16] described the valence band through semi-analytical non-parabolic and anisotropic description calibrated to *k.p*. Our work, shows that the valence band problem is exceedingly more complicated in the case of nanowires. Here, enhanced coupling due to structural quantization, and large sensitivity of the band shape/curvature to potential variations in the lattice, especially in the [100] wire orientation, make the effective mass approximation (EMA) and the *k.p* method insufficient, and call for atomistic treatment [17, 18, 19]. (This is in contrast to the variations of the curvature of the conduction bands which show much less sensitivity to lattice potential variations). Atomistic modeling can automatically capture these effects and provide the dispersion of the structure under consideration. In addition, nanowire cross sectional shapes, quantization surfaces, transport orientations, and arbitrary crystal distortions can be automatically treated in an atomistic model.

*Summary of results:* The anisotropy of the Si heavy-hole valence band strongly affects the preference of charge placement in the wires' cross section and differs for different oriented wires (Fig. 1b). The heavy quantization masses in the [110] and [112] orientations cause preferable charge accumulation on the (110) and (112) surfaces rather than the (100) surfaces controlled by a lighter quantization mass. The semiconductor capacitance $C_S$, as also in the case of NMOS nanowires [11], is important and can degrade the capacitance of the device by up to ~30% for wires in all transport orientations. In terms of ON-current capabilities, p-type nanowire transport will be preferable in the [111] oriented devices that have the largest carrier velocities closely followed by the [110] devices. [100] nanowires indicate much lower performance due to the enhanced band warping and curvature variations in their dispersions.

*This paper is organized as follows:* In section 2 the simulation approach is highlighted. Section 3, contains the numerical results. Sections 3(a,b,c) examine the [100], [110] and [111] oriented 3nm and 6nm square wires' dispersion and charge distribution under potential variations in the lattice. Section 3(c) examines the performance of the nanowires in terms of total gate capacitance, quantum capacitance, injection velocity, and drive current capabilities. Finally, in section 4 explanations about the behavior of nanowires in different orientations are provided from features of the *bulk* bandstructure of Si.

## 2. Approach

The technical approach used in this work, as well as the justification for the validity of our model has been extensively described in reference [11]. We repeat Fig. 1 here as a simple overview, but the reader should refer to [11] for details. In summary, the computational model consists of three steps (Fig. 1a) as follows:

1. The bandstructure of the wire is calculated using the atomistic tight-binding model. The atoms that reside on the surface of the nanowire are passivated in the sp$^3$ hybridization scheme [20].
2. A semiclassical top-of-the-barrier ballistic model is used to fill the dispersion states and compute the transport characteristics [12, 13].
3. A 2D Poisson equation is solved in the cross section of the wire to obtain the electrostatic potential. The electrostatic potential is added to the diagonal on-site elements of the atomistic Hamiltonian as an effective potential for recalculating the bandstructure until self consistency is achieved.

Although the transport model used is simplistic, it allows for examining how the bandstructure of the nanowire alone will affect its ballistic transport characteristics. Extensions to true 3D transport treatment with interface roughness or alloy disorder can be achieved with enough computational power [21]. The same conclusions to parts of this work can be similarly obtained, but the simple model used here provides critical physical insight. It is however, the simplicity of the transport model, which allows to shed light on the importance of the dispersion details, which might get lost otherwise in a full-fledged quantum transport simulation.

## 3. Results

In this section, an analysis of the transport properties of 3nm and 6nm square nanowires in [100], [110] and [111] orientations is performed. In all cases, rectangular nanowires with oxide thickness $t_{ox}$=1.1nm are assumed. Strong interactions between the nanowire valence bands, as well as potential variations introduce non-trivial features in the dispersions. Their dispersion properties (in terms of carrier velocities, and quantum capacitance) and the shape of the charge distribution in the cross section of the each wire are analyzed. The relative performances are then compared.

*(a) Hole transport in [100] oriented nanowires.*

*The 3nm [100] wire:*

Figure 2(a-d) shows device features for a 3nm square [100] oriented nanowire under low ($V_G$=0V) and high ($V_G$=1.2V) gate biases. (The drain bias used is $V_D$=0.5V in all cases throughout this work). Under low gate biases, the lattice is almost empty of charge (Fig. 2a) and the dispersion relation (Fig. 2b) is almost the empty lattice equilibrium dispersion. The dispersion is a mixture of heavy-, light-hole, and split-off bands, indicating strong band coupling and warping. Under high biases, there is significant charge filling of the lattice as shown in Fig. 2c. Figure 2d shows the dispersion of the nanowire under high gate bias. Similar to the case of electron conduction band transport [11], charge filling of the lattice causes changes in the dispersion of the nanowire even at the 3nm wire length scale (Fig. 2d). The nature of the changes in the dispersions is not a simple shift in the bands' position, but rather strong warping, splittings and change of shape/curvature. Qualitative explanations on the shape of the dispersion will be given in section 4 of this paper.

*The 6nm [100] case:*

Figure 2e-h shows for a 6nm square wire the same characteristics as shown for the 3nm wire of Fig. 2a-d. Under low bias conditions, the charge shown in Fig. 2e is also placed in the middle of the channel. As the bias increases, the charge shifts towards the corners of the device as shown in Fig. 2g. Similar corner effects in 6nm wires have also been calculated for the case of NMOS nanowire devices [22]. The corner effect is an electrostatic one, since the corners can reach inversion faster than the rest of the surfaces. Significant changes are also observed in the dispersion of the charge filled lattice (Fig. 2h) compared to that of the empty one (Fig. 2f).

*(b) Hole transport in [110] oriented nanowires.*

*The 3nm [110] case:*

Fig. 3a-d shows the same quantities for the 3nm [110] oriented wire, as shown in Fig. 2a-d for the 3nm [100] oriented wire. The [110] wire's valence band dispersion shown in Fig. 3b is significantly different than that of the [100] wire. Its' highest energy valley can be roughly described by effective mass, estimated to be ~$0.14m_0$. (This is done using a parabolic dispersion constructed to fit the dispersion up to 0.1eV). The change in the dispersion under potential variations is also evident in this wire as shown in Fig. 3d. An estimate of the effective mass of the highest energy valley of this dispersion is calculated to be ~$0.17m_0$.

An obvious difference between the [110] and the [100] wire is observed in the charge distribution in the cross section of the wire. As clearly observed in Fig. 3a, under low bias the charge distribution is centered in the middle of the channel, preferably along its vertical *y*-axis along the [1-10] direction rather than the *x*-axis along [001]. This is an effect resulting from the anisotropy of the heavy-hole quantization mass in the [1-10] and [001] directions. Quantization in the $\langle 100 \rangle$ equivalent direction is subject to a lighter mass, and the wavefunction is shifted farther away from the Si/SiO$_2$ interface. The quantization mass in $\langle 110 \rangle$ is heavier, forcing the wavefunction more towards the (1-10) surface (see Fig. 3a). This is even more evident under high biases as shown in Fig. 3c. Although the potential distribution in the cross section of the wire (inset of Fig. 3d) is symmetric, the charge distribution in the cross section is preferably accumulated on the (1-10) surface (top/bottom) rather than the (001) surface (left/right). More explanations on this will be provided in section 4. It is mentioned here, however, that the light-hole and split-off bands are almost isotropic, and do not have significant contributions to the overall anisotropic behavior, despite the band mixing that takes place.

*The 6nm [110] case:*

Figure 3e-h shows the same features for the 6nm x 6nm [110] wire as Fig. 3a-d for a 3nm x 3nm device. Under low bias conditions (Fig. 3e), the charge distribution still shows the preferential distribution along the *y*-[1-10] axis as in Fig. 3a. The low bias dispersion at energies close to the valence band edge indicates light masses that, however, get heavier for lower energies (Fig. 3f) through changes in their curvature. At high biases, as shown in Fig. 3g, the charge distribution shifts towards the corners of the device. Unlike the 6nm [100] wire case in Fig. 2g, however, the charge is closer to the interface, and preferably accumulates on the top/bottom (1-10) surfaces rather than the left/right (001) surfaces. Comparing the charge distribution in Fig. 3g with that of Fig. 3c, it appears that the inversion lobes that form along the (1-10) surface of Fig. 3c are now on the four corners of Fig. 3g. It is reasonable to assume that the 6nm wire behaves as two 3nm wires, with two inversion lobes in the left/right each. (Looking at the dispersion of Fig. 3h, a second pair of light mass subbands above the Fermi level now appears as compared to the dispersion of the 3nm wire in Fig. 3d).

*(c) Hole transport in [111] oriented nanowires.*

*The 3nm [111] case:*

In Fig 4, the same analysis is performed for the [111] oriented wire. The [111] wire's valence band of Fig. 4b is again different than that of the previous two described wires. Its highest valley can be roughly described by an approximate effective mass, estimated to be ~0.13$m_0$. Under high bias, the dispersion changes (Fig. 4d)), and now the effective mass of the highest energy valley becomes ~0.11$m_0$. Under low bias, the charge distribution in Fig. 4a is centered in the middle of the channel. Under high bias (Fig. 4c), the charge is still mostly symmetric around the cross section. This originates from the similarity in the dispersions along the quantization surfaces of this wire which are the (1-10) and the (11-2) surfaces, although a small degree of asymmetry can be observed. More details on this will be provided in section 4.

*The 6nm [111] case:*

Fig 4e-h again shows the same analysis for the 6nm square [111] wire. Under low bias conditions, the charge distribution is almost symmetric in the center of the wire as shown in Fig. 4e. The dispersions' curvature at energies close to the valence band edge is large, indicating light masses. As the energy increases, however, the curvature reduces, and the bands become heavier (Fig. 4f). At high biases, the charge accumulates close and along the surfaces, slightly preferable along the (1-10) top/bottom surfaces rather than the (11-2) left/right surfaces. The quantization masses of (1-10) and (11-2) surfaces are similar, which makes the charge distribution more symmetric. They are heavier than the

(100) quantization masses, which allows the charge placement closer to the interfaces (comparing at the same gate overdrive). The high bias dispersion of this wire in Fig. 4h, as also observed in the case of the [110] wire, has light dispersion features for energies above the Fermi level, similar to the light bands in the low bias case for the 3nm wire of Fig. 4d. Just by looking at the dispersions, the 6nm wire under inversion behaves similarly to the 3nm wire, with a larger number of subbands.

*Implications of the charge placement to device performance:*

The different quantization behavior of different surfaces should be taken into serious consideration when evaluating multi-surface type of devices. Kobayashi *et al.* in [5], showed experimentally that different quantizations impact the performance of nanowire devices through $V_T$ fluctuations and ON-current variations. In that work, it was shown that $V_T$ and $I_{ON}$ of PMOS nanowire devices are very sensitive to [100] side variations, but much less sensitive to [110] side variations. This is evidence of the heavier [110] mass quantization that does not allow large subband variations with size fluctuations, and enforces a thinner inversion charge layer on the surface. On the other hand, placement of the charge very close to the interface, might make the device more susceptible to surface roughness scattering. These issues should be taken into consideration in the design of nanowire devices.

*(b) Device performance comparison of NWs in different orientations*

Figure 5 shows a performance comparison between the wires in the [100], [110] and [111] orientations for the 3nm and 6nm square nanowires. The various performance quantities shown further on, are all compared at the same OFF current ($I_{OFF}$) for all devices.

*Gating induces the same capacitance / charge in all wire directions:* Figure 5a shows the total gate capacitance ($C_G$) vs. gate bias ($V_G$) of the three wires at the same $I_{OFF}$. The total capacitance in the three wires is very similar for all gate biases, as also observed in the case of the n-type nanowires [11]. This is an indication that the same amount of inversion charge is accumulated in all wires irrespective of their orientations.

*Low semiconductor capacitance ($C_S$) degrades the gate capacitance ($C_G$) from the oxide capacitance ($C_{OX}$) by ~30%:* This amount corresponds to an effective increase in the oxide thickness of 0.35nm, ( ~30% of the physical gate oxide thickness of $t_{ox}$=1.1nm). From $C_G = C_S C_{OX}/(C_S + C_{OX})$, with $C_G$=0.35nF/m (maximum value of Fig. 5a), and $C_{OX}$=0.483 nF/m [23], $C_S$ can is calculated to be $C_S = 1.27 nF/m$, which is only ~3 larger than the value of the oxide capacitance. Comparing the $C_S$ value to the quantum capacitance of the nanowires shown in Fig. 5b, ($C_Q$<3nF/nm for all wire orientations), the $C_S$ is reduced from $C_Q$ by a factor of ~2. This situation is the same in the case of the conduction band transport properties, and has also been observed in thin body devices and even bulk MOSFETs [24]. As discussed by Pal, this happens whenever the potential well formed in the channel is bias dependent. Equivalently, the charge centroid is placed further away from the interface, and $C_S$ is reduced. $C_Q$ being very similar for all nanowires, on the other hand, results in very similar final gate capacitances for all wires as well.

*Velocity controls the transport differences in different orientated wires:* Figure 5c shows the injection velocities of the wires vs. gate bias ($V_G$). The [111] wire has the largest velocities, followed by the [110] wire, whereas the [100] wire has the lowest velocities, as would be expected from the subband masses estimated earlier for the [110] and [111] wires. (In 1D, under the parabolic band approximation, the velocity is proportional to $v \sim 1/\sqrt{m^*}$). The lowest velocity in the [100] wire, is a result of the enhanced warping in its dispersion, which slows carriers down. The injection velocities increase as high energy carrier states are being populated. This increase in velocity with gate bias can reach up to 40% in the [100] wires and even up to 29% in the [111] and 21% in the [110] wire orientation cases.

*Velocity differences affect the I-V differences:* Since the charge is the same in all devices, the velocity difference directly reflects on the ballistic $I_{DS}$ as shown in Fig. 5d in which the drive current capabilities of the wires are compared at the same $I_{OFF}$. The [111] and [110] wires perform better than the [100] wire in terms of ON-current capabilities. Comparing to the [100] wire, the [110] wire can transport almost 2 times as much current, whereas the [111] wire almost 2.5 times higher current.

*Transport characteristics for the 6nm wires:* The same analysis has been performed in Fig. 5e-h for the 6nm x 6nm wires. Similar observations as in the 3nm wire case are concluded for the total capacitance (Fig. 5e), the quantum capacitance (Fig. 5f), the velocities (Fig. 5g), and the relative $I_{DS}$ performance between the different orientation wires. The relative differences in the performance are slightly different than in the 3nm case, however, qualitatively the same conclusions can be drawn. Comparing to the 3nm wire case, the gate capacitance and the quantum capacitance doubles for the 6nm wire. The velocity is lower than that in the 3nm wire cases because slower carrier velocity subbands are now been occupied. Finally, the current is increased by ~60% (comparing at the same gate overdrive) when going from the 3nm x 3nm wire to the 6nm x 6nm wire (while the area is increased by 4 times).

Although we do not have a comparison of our results to experimental data because at the moment we could not identify a comprehensive experimental study for the orientation dependence of ballistic Si nanowire devices, we would like to mention that the $sp^3d^5s^*$-SO tight-binding model was used to verify several experimental observations both for the orientation behavior of the transport properties of quantum wells, as well as several other type of electronic properties. In [11] we present a list of all the different experimental observations that the model was used to verify, and we believe the results we present here accurately project on the orientation behavior of scaled nanowires. In addition to that, this work shows that for a device operating in the quantum capacitance regime, variations in the density of states of the channel do not affect the total capacitance of the device. In order to improve the drive current capabilities of such devices, one can improve the injection velocity (by reducing $m^*$) in the expense of the density of states, since the total capacitance might not suffer extensively.

## 4. Discussion

*The k-space energy surface under different orientations-quantization behavior:*
In this section, intuitive explanations to the quantization behavior and the shape of the transport dispersions in different transport orientations are provided. The study of the curvatures and masses of the surface slices through the 3D bulk *E(k)* in different directions can guide towards understanding of the nanowire quantization behavior. The main reasoning comes from the anisotropy of the heavy-hole bandstructure, which defines the transport and quantization masses in different orientations. The heavier the quantization mass of a surface, the closer the charge will accumulate near the surface under inversion conditions. The heavy-hole (100), (110), (112) and (111) surfaces are investigated in Fig. 6. (The light-hole and split-off bands are almost isotropic and do not enforce strong preferable charge distribution on any surface). Therefore, although band coupling takes place in nanowires, as we have observed from the dispersions of Fig. 2, 3, and 4, most of the anisotropic behavior results from the heavy-hole. The intuitive explanations we provide below are based on a single HH band. Qualitatively, they can provide explanations for the orientation behavior, and guidance towards the design of nanowire transistors).

*The (100) E(k) surface:* Figure 6a shows the 2D $E(k_{[100]}, k_{[010]})$ of the (100) surface. The contours at *E=0.2eV* and *E=1eV* are plotted. The bandstructure is anisotropic, with heavier mass dispersion along the [110] direction ($m^*_{[110]}$= -0.581$m_0$) and lighter along the [100] direction ($m^*_{[100]}$= -0.276$m_0$). Quantization in [110] will utilize heavier quantization masses, than quantization in [100] directions. At the same inversion conditions in a nanowire, the charge distribution will preferably reside closer to the (110) surface than the (100) surface, which is exactly what observed in Fig. 3d and 3g. On the other hand, the symmetry of the heavy hole in the [001] and [010] directions, results in the symmetric charge distribution in the [100] oriented wires of Fig. 2. Of course, the [100] wire can be quantized in the [110] and [1-10] directions. The charge distribution in this case will reside closer to the interface than what is shown in Fig. 2c, g.

*The (110) E(k) surface:* The differences in the quantization behavior in the [110] directed wire (with (100) and (110) equivalent quantization surfaces) is more evident in the $E(k_{[1-10]}, k_{[001]})$ shown in Fig. 6b. This energy contour is a cross section along the 45° diagonal line of Fig. 6a. The [010] and [10-1] orientations indicated, are the quantization directions of the [110] oriented wire.

*The (111) E(k) surface:* The quantization of the [111] channels described earlier is determined from the (111) $E(k)$ surface, with [1-10] and [11-2] quantized sides, shown in Fig. 6c. The energy surface does not look very different in the two directions, although minor details can be found. The charge distribution of Fig. 4 is due to this reason almost symmetric in the two quantization directions (with some minor differences).

*(b) Understanding the transport dispersion features:*

*Relevant band extraction method:* The bulk energy contour surfaces can provide indications for the shape of the dispersions in different orientations (either quantum wells or nanowires). Under any physical quantization, the relative energy/subband levels will follow the "particle in a box" quantization, and move away from the center of the band minima at a rate of $k_n=n.\pi/L$, where $L$ is the quantization size. The smaller the physical domain, the larger the corresponding quantized $k$. For example, the relevant energy dispersion for a *quantum well* quantized in the [010]-direction, will be the horizontal energy surfaces in the [100]-[001] plane, passing through the lines drawn in Fig. 6(e). Similarly, the relevant energy dispersions in [110] transport direction with quantization of the (1-10) surface are given by surfaces drawn through the 45° lines shown in Fig. 6(e). (The relevant dispersions in the [111] transport direction with quantization of the (1-10) surface, will be surfaces perpendicular to the 45° lines of the (112) surface of Fig. 6(d) along [111]). Figures 6f-h show the transport direction subbands for the [100], [110] and [111] orientations as the (010), (1-10) and (1-10) respectively are quantized, and for the $k$-vector in the remaining quantization direction set to $k=0$ (i.e. the quantization lines in Fig. 6e). The quantization assumed is $L$=6nm (equivalently 12, 17 and 7 unit cells in the [110], [110] and [111] wires respectively).

*Band shape in [100]:* Figure 6(f) shows energy subbands of the (010)/[100] structure. These will be relavant subbands in [100] orientation and partially explain the oscillating behavior observed in the subbands of the [100] wire dispersions of Fig. 2.

(This is just an indication on where these subbands originate from. The method for an exact reconstruction of the nanowire subbands from the bulk bandstructure will be presented in the next section). Strong band coupling however, allows only qualitative understanding for the subband form, in contrast to the case of n-type wells, for which a good agreement can be achieved between this semi-analytical method and the full TB dispersions.

*Band shape in [110]:* The (110)/[110] direction subbands are drawn in Fig. 6(g). (45° lines in Fig. 6(e)). The subband passing through the center of the energy contour is of heavy-mass, however, as the structure is quantized in the [110] direction, the subbands shift away from the center and become lighter. This explains partially the shape of the subbands in the [110] wires of Fig. 3(c), where the 3nm quantized wire has light mass subbands, whereas the 6nm quantized wire in Fig. 3(g) has heavier subbands. Of course under electrostatic quantization in Fig. 3(h), the highest energy subbands become lighter again (similar to the case of structural quantization).

*Band shape in [111]:* The [111] case, is similar to the [110] case as shown in the subbands of Fig. 6(h). Here, the subbands are drawn by taking lines at 45° in Fig. 6(d) along [111] perpendicular to the [1-10] direction. Lines passing from the center (weak quantization), are heavy subbands, but as the structure is more quantized the subbands become lighter. The subband features in this case shine light into the [111] nanowire dispersion features of Fig. 4(b) under strong quantization, and Fig. 4(f) under weak quantization. Strongly quantized subbands will look more like the large *n*-value subbands (with lighter masses), while weakly quantized subbands will look like the smaller *n*-value subbands (with heavier masses). Electrostatic quantization will have a similar effect as physical quantization in the curvature of the subbands, transforming the equilibrium dispersion of the 6nm wire (Fig. 4f) into a dispersion with subbands similar to the 3nm wire subbands.

*(c) Semi-analytical construction of the NW dispersion from the bulk E(k):*

The study of the surface slices through the bulk *E(k)* in different directions can guide some understanding of the nanowire dispersions. After picking a surface in a particular direction, the two remaining *k*-directions are quantized due to the lateral nanowire confinement. The quantized point on the plane corresponds to a single *k*-point in the transport direction dispersion. To obtain all *k*-points of the 1D subband in the transport direction, the surface slice needs to be translated further into the bulk dispersion and the lateral quantization must be redone. All relevant 1D subbands from the bulk 3D *E(k)* along the transport direction are obtained by shifting in *k*-space in both two relevant quantization directions $k_A=n.\pi/L_A$ and $k_B=m.\pi/L_B$, where $k_A$, $k_B$ and $L_A$, $L_B$ are the two quantization *k*-space directions and quantization lengths, for all *n, m*.

Figure 7 compares the full numerical dispersions to the semi-analytical dispersions constructed from the *bulk E(k)* for different wire cross sections (3nm, 12nm) and wire directions [100] and [110]. All heavy-hole, light-hole and split-off bands are included in this construction. In the cases of the 12nm wires, the envelope of the bands in the semi-analytical dispersion (Fig. 7 c, d) in both the [100] and [110] cases approximates the actual nanowire dispersion (Fig. 7 a, b) very closely. Enhanced band interactions in the case of the dispersions of the actual wires create qualitative differences in these multi-band dispersion figures. Potential variations are however not part of this analysis, which can have large impact on the dispersions in the actual wire. In the case of the smaller (3nm) wires, both in [100] and [110] orientations, some of the trends and band shapes of the actual dispersion (Fig. 7 e, f) are captured by the semi-analytical construction (Fig. 7 g, h), however due to the enhanced coupling, only small qualitative similarities can be observed.

The semi-analytical construction method can only provide qualitative indications as of the subband form of the different oriented nanowires. Band interactions and the effect of potential variations is not included in this approach, however, reasonable understanding about the subbands of nanowires can be extracted from the simple bulk bandstructure. It is noted here, that in the case of the conduction band dispersion, or valence band of 2D quantum wells, where the band coupling is reduced, the semi-analytical construction method gives a much closer result to the actual dispersion of the structure [11, 25].

## 5. Conclusions

Transport properties of nanowires in different transport orientations ([100], [110] and [111]) are examined using a 20 orbital sp$^3$d$^5$s* spin-orbit-coupled atomistic TB model self-consistently coupled to a 2D Poisson solver. A semiclassical ballistic model is used to calculate the current-voltage characteristics of the nanowires. The dispersions of the nanowires cannot be described within the EMA because of the enhanced band coupling that induces large warping in the dispersions, especially for the [100] oriented wires. In addition, the dispersion shapes are strongly bias dependent.

In the ballistic limit, the [111] wire has the largest carrier velocities and ON-current capabilities, followed by the [110] wire. The [100] wire is the worst in terms of both carrier velocities and drive current capabilities. The semiconductor capacitance ($C_S$) is important for nanowire devices and degrades the gate capacitance by ~30%. This effect is very similar for all wire orientations, for both 3nm and 6nm square wires.

The shape of the charge distribution in the cross section of the different oriented wires differs for each wire according to the quantization mass that each surface feels. The [100] wire examined with (010)/(001) quantization surfaces has a symmetric charge distribution in its cross section. In the case of [110] wires, with (1-10)/(100) quantization surfaces, the charge preferably accumulates on the (1-10) surface due to its largest quantization mass. The [111] wire, with similar quantization mass in the (1-10)/(11-2) surfaces has only slightly non-symmetric charge distribution profile on the two surfaces. These observations can provide guidance towards the design of multi surface devices such as nanowires and FinFETs.

Finally, the authors would like to mention that the simulator used in this study will be released as an enhanced version of the Bandstructure Lab on nanoHUB.org [26]. This simulation engine will allow any user to duplicate the simulation results presented here. Over 1,800 users have run over 12,000 simulations in the existing Bandstructure Lab, which has not yet included the charge self-consistent transport model we demonstrate here. This new charge self-consistent capability has been added very recently (August 2008).

# 6. Acknowledgements

This work was funded by the Semiconductor Research Corporation (SRC). The computational resources for this work were provided through nanoHUB.org by the Network for Computational Nanotechnology (NCN). The authors would like to acknowledge Prof. Timothy Boykin of University of Alabama at Huntsville for tight-binding discussions.

Figure Captions

Figure 1:

(a) Simulation procedure schematic. Using an atomistic sp$^3$d$^5$s*-SO tight-binding model, the bandstructure of the nanowire under consideration is calculated. A semiclassical ballistic model is then used to calculate the charge distribution in the wire from the source and drain Fermi levels. The charge is used in a 2D Poisson for the electrostatic solution of the potential in the cross section of the wire. The oxide thickness assumed is $t_{ox}$=1.1nm. The whole process is done self consistently. (b) The lattice in the wire transport orientations (surfaces) used – [100], [110] and [111]. The charge distribution in the back is distinctively different for the three cases.

Figure 2:

Device features for a [100] rectangular wire. (a-d) The 3nm square wire. (a-b) The 2D cross section showing the charge distribution and dispersion *E(k)* under $V_G$=0V gate bias conditions. $a_0$ is the wires' unit cell length. (c-d) The 2D cross section showing the charge distribution and dispersion *E(k)* under $V_G$=1.2V, high gate bias conditions. The dots indicate the underlying atomic positions. $E_{fs}$ is the source Fermi level. (Zero energy indicates the conduction subband edge). (e-h) Same features as in (a-d) for the 6nm square wire.

Figure 3:

Device features for a [110] rectangular wire. (a-d) The 3nm square wire. (a-b) The 2D cross section showing the charge distribution and dispersion *E(k)* under $V_G$=0V gate bias conditions. $a_0'$ is the wires' unit cell length. (c-d) The 2D cross section showing the charge distribution and dispersion *E(k)* under $V_G$=1.2V, high gate bias conditions. Inset of (d): The electrostatic potential contour in the cross section of the nanowire. The dots indicate the underlying atomic positions. $E_{fs}$ is the source Fermi level. (Zero energy indicates the conduction subband edge). The masses indicated are approximate effective masses for the highest energy level subband. (e-h) Same features as in (a-d) for the 6nm square wire.

Figure 4:

Device features for a [111] rectangular wire. (a-d) The 3nm square wire. (a-b) The 2D cross section showing the charge distribution and dispersion $E(k)$ under $V_G$=0V gate bias conditions. $a_0''$ is the wires' unit cell length. (c-d) The 2D cross section showing the charge distribution and dispersion $E(k)$ under $V_G$=1.2V, high gate bias conditions. The dots indicate the underlying atomic positions. $E_{fs}$ is the source Fermi level. (Zero energy indicates the conduction subband edge). The masses indicated are approximate effective masses for the highest energy level subband. (e-h) Same features as in (a-d) for the 6nm square wire.

Figure 5:

Performance comparison of the square wires in the [100], [110] and [111] directions at the same OFF-current ($I_{OFF}$). (a-d) The 3nm square wire: (a) The gate capacitance $C_G$ vs. gate bias ($V_G$). The capacitance is similar for all wires, and degraded from the oxide capacitance by an amount that corresponds to an increase in the effective oxide thickness of 0.35nm. (b) The quantum capacitance $C_Q$ vs. $V_G$ of the three devices, which is a measure of the density of states at the Fermi level. (c) Comparison between the injection velocities of the nanowires vs. $V_G$. In all cases, the velocity is not constant, but increases as the gate bias increases. The increase is calculated by the difference between the value at high $V_G$ and the value at low $V_G$. (d) The $I_{DS}$ vs. $V_G$ for the three wires at the same $I_{OFF}$. The velocity difference directly reflects on the current differences. (e-h) The 6nm square wire: Same features as in (a-d) respectively.

Figure 6:

(a-d) Energy surface contours of the heavy hole calculated using the full 3D $k$-space information of the Si Brillouin zone. The energy contours for $E$=-0.2eV and $E$=-1eV are plotted. (a) The (100) surface. The anisotropy is evident in the [100] and [110] directions. (b) The (110) surface. (Or equivalently, 45° "cut" through the center of (a) into the surface. (c) The (111) surface. (d) The (112) surface. The [1-10] and [111] in plane directions are indicated. (e-h) Extraction of relevant quantization subbands. (e) The (100) energy surface with a few relevant quantization lines under quantization in the (010) (vertical $k_{[100]}$=constant lines) and (1-10) (45° lines) surfaces. Under quantization in (010), the shift in the $k$-value is given by $\Delta k = n.\pi/L$, where $n$ is the subband index, and $L$ is the quantization length (6nm in this case) (f) Relevant subbands in (010) surface quantization and [100] transport. (g) Relevant subbands in (1-10) surface quantization and [110] transport. (h) Relevant subbands in (1-10) surface quantization and [111] transport.

Figure 7:

(a-b, e-f) The $E(k)$ dispersions for square nanowires calculated using the full numerical TB model. (a) 12nm [100] wire. (b) 12nm [110] wire. (e) 3nm [100] wire. (f) 3nm [110] wire. (c-d, g-h) The $E(k)$ dispersions for square nanowires calculated using the semi-analytical construction from the bulk bandstructure. (c) 12nm [100] wire. (d) 12nm [110] wire. (g) 3nm [100] wire. (h) 3nm [110] wire.

Figure 1: The model

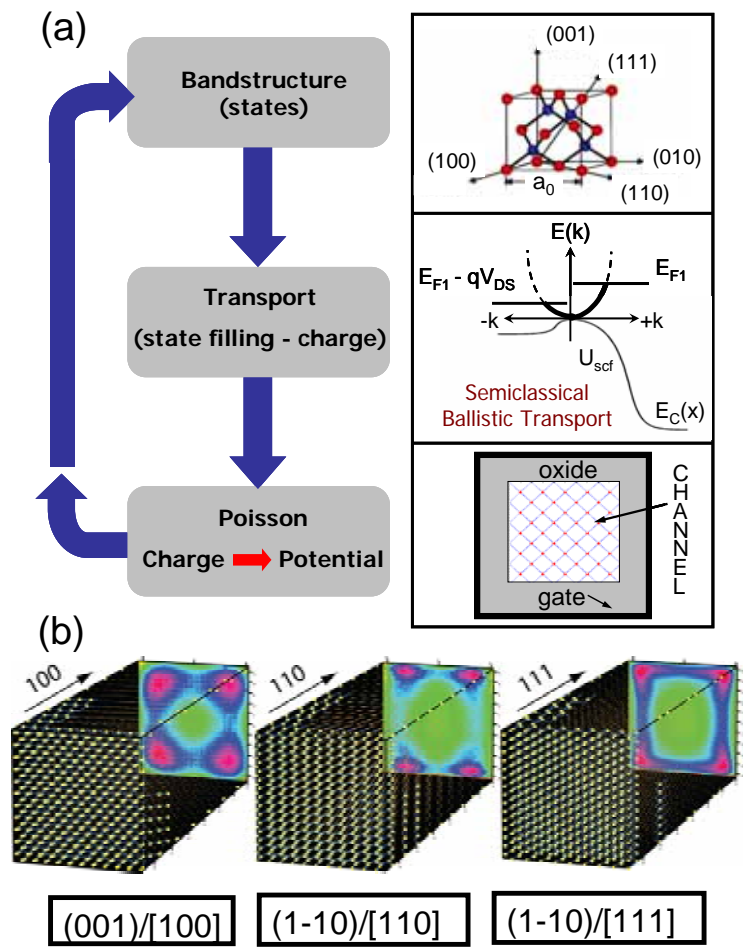

Figure 2: [100] wire

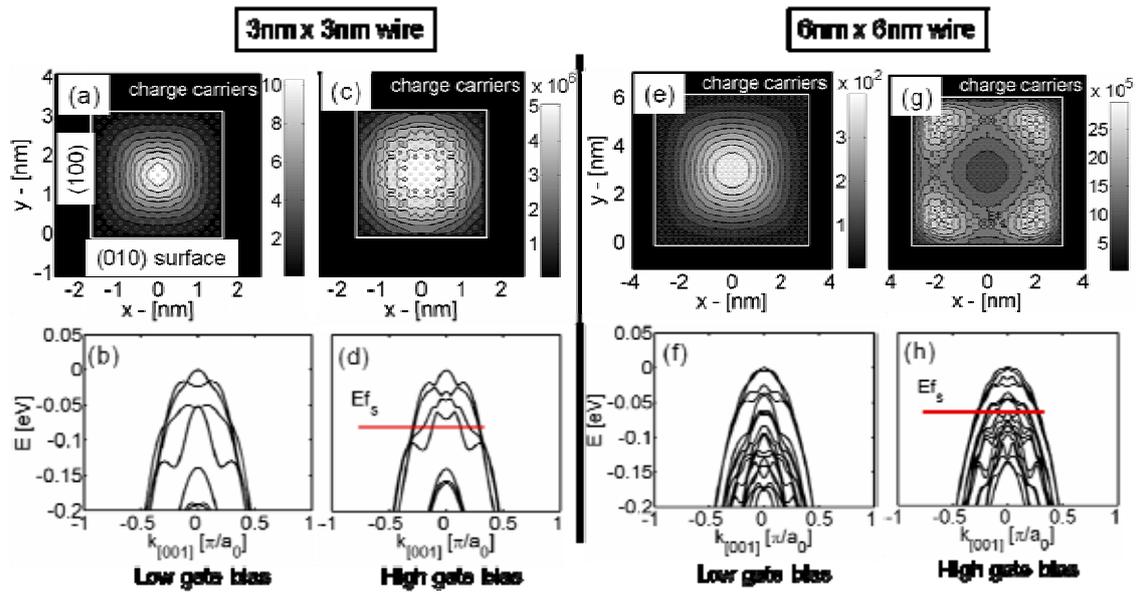

Figure 3: [110] wire

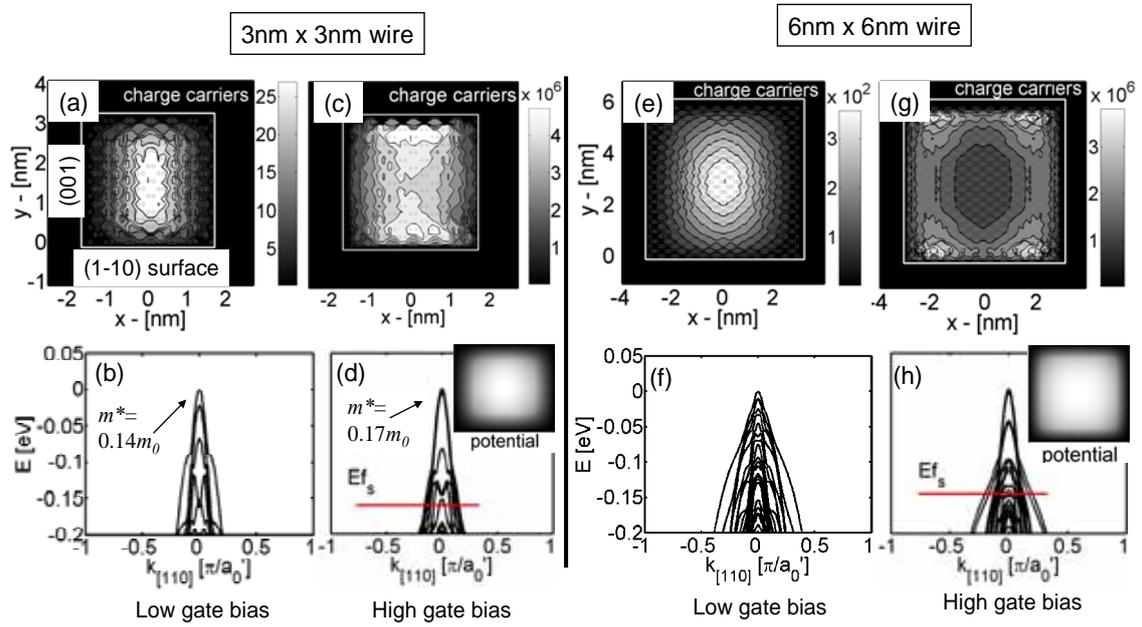

# Figure 4: [111] wire

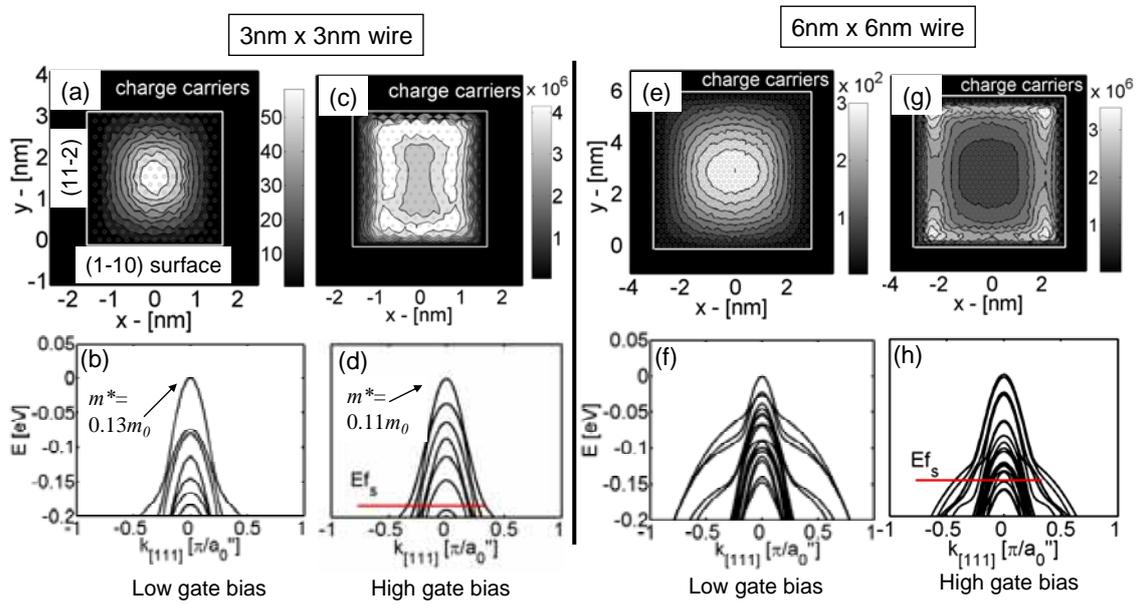

Figure 5: ID-VG

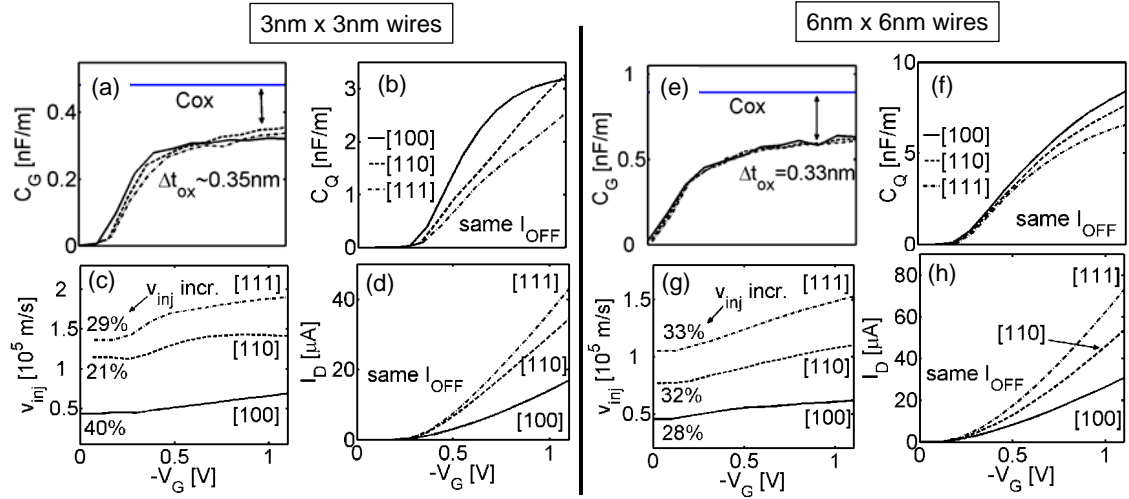

Figure 6: quantization surfaces and subbands

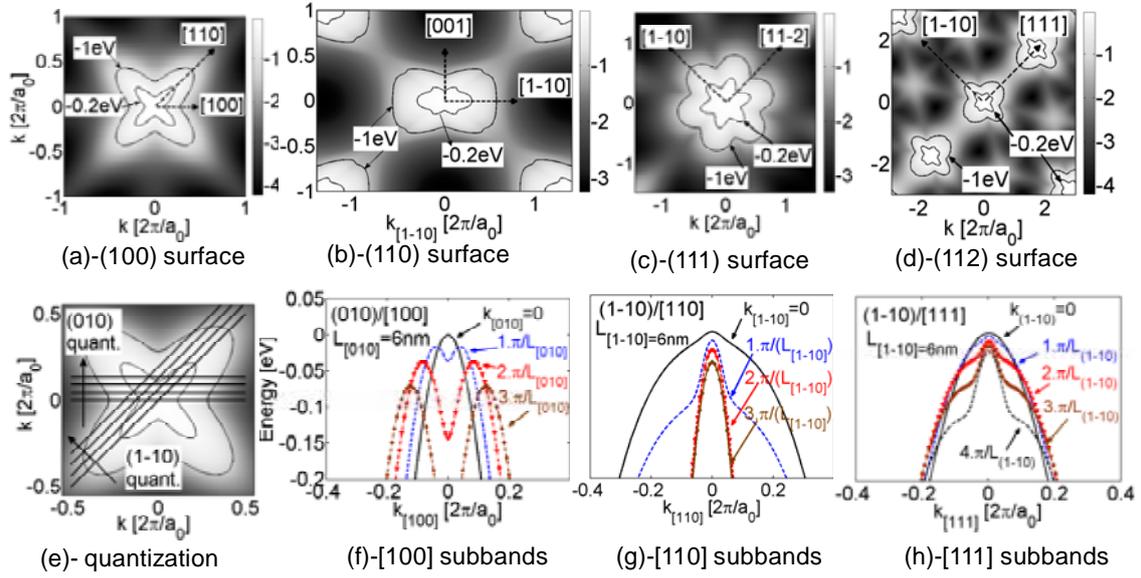

Figure 7: Semi-analytical construction of the NW dispersion

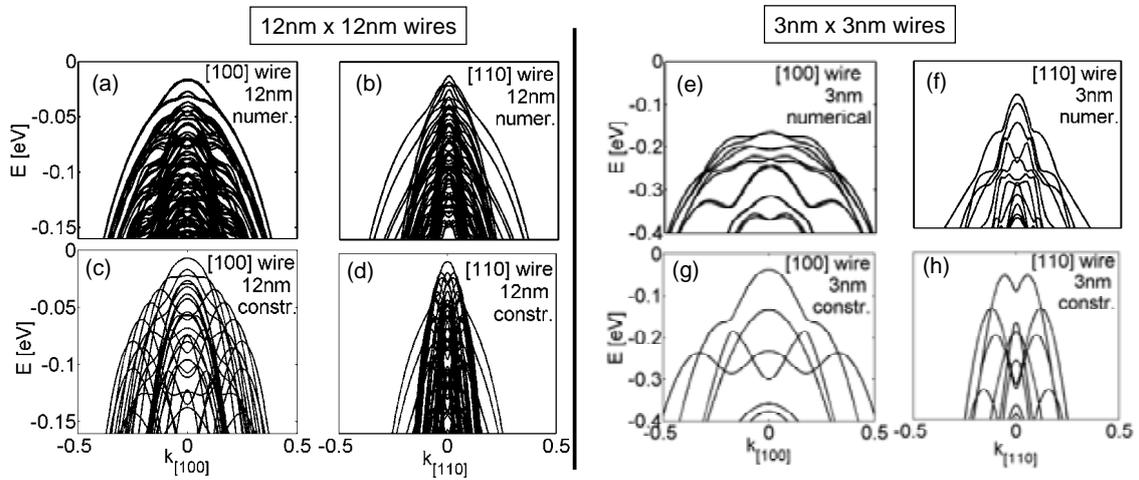